\documentclass[preprint2,twocolumn,times,tighten]{aastex631}

\usepackage{graphicx}	
\usepackage{amsmath}	
\usepackage{multirow}

\begin{document}

\title{Probing Three-Dimensional Magnetic Fields: IV - Synchrotron Polarization Derivative and Vision Transformer}

\author[0000-0002-8455-0805]{Yue Hu*}
\affiliation{Institute for Advanced Study, 1 Einstein Drive, Princeton, NJ 08540, USA}

\author{Alex Lazarian}
\affiliation{Department of Astronomy, University of Wisconsin-Madison, Madison, WI 53706, USA}

\email{yuehu@ias.edu, *NASA Hubble Fellow}
\email{lazarian@astro.wisc.edu}



\begin{abstract}
Measuring the 3D spatial distribution of magnetic fields in the interstellar medium and the intracluster medium is crucial yet challenging. The probing of 3D magnetic field's 3D distribution, including the field plane-of-sky orientation ($\psi$), the magnetic field's inclination angle ($\gamma$) relative to the line of sight, and magnetization ($\sim$ the inverse Alfv\'en Mach number $M_A^{-1}$), at different distances from the observer makes the task even more formidable. However, the anisotropy and Faraday decorrelation effect in polarized synchrotron emission offer a unique solution. We show that due to the Faraday decorrelation, only regions up to a certain effective path length along the line of sight contribute to the statistical correlation of the measured polarization. The 3D spatial information can be consequently derived from synchrotron polarization derivatives (SPDs), which are calculated from the difference in synchrotron polarization across two wavelengths. We find that the 3D magnetic field can be estimated from the anisotropy observed in SPD: the elongation direction of the SPD structures probes $\psi$ and the degree of SPD anisotropy, along with its morphological curvature, provides insights into $M_A^{-1}$ and $\gamma$. To extract these anisotropic features and their correlation with the 3D magnetic field, we propose utilizing a machine learning approach, specifically the Vision Transformer (ViT) architecture, which was exemplified by the success of the ChatGPT. We train the ViT using synthetic synchrotron observations generated from MHD turbulence simulations in sub-Alfv\'enic and super-Alfv\'enic conditions. We show that ViT's application to multi-wavelength SPDs can successfully reconstruct the 3D magnetic fields' 3D spatial distribution.
\end{abstract}



\section{Introduction} \label{sec:intro}
Polarized synchrotron emission is a fundamental tool for investigating magnetic fields across diverse astrophysical environments, including the interstellar medium (ISM; \citealt{1983Natur.304..243M,2008A&A...482..783X,2009A&A...503..827X,2012SSRv..166..231R}), the circumgalactic medium (CGM; \citealt{2001SSRv...99..243B,2015A&ARv..24....4B}), and the intracluster medium (ICM; \citealt{2004IJMPD..13.1549G,2014IJMPD..2330007B,2021MNRAS.502.2518S,2024NatCo..15.1006H}). It provides essential insights into cosmic ray physics \citep{1966ApJ...146..480J,1978MNRAS.182..443B,2012SSRv..166...71B,2014ApJ...783...91C,2014ApJ...785....1B,2022ApJ...925...48X,2022ApJ...934..136X}, galactic dynamics \citep{2001SSRv...99..243B,2018MNRAS.476..235R,2022ApJ...941...92H,2023MNRAS.519.1068L,2023ApJ...952....4B}, and galaxy cluster evolution \citep{2004IJMPD..13.1549G,2011ApJ...743...16Z,2014IJMPD..2330007B,2022A&A...657A..56K}. Given its broad application and crucial importance, getting more detailed information from synchrotron polarization is extremely important.

Synchrotron polarization is predominantly used to trace the plane-of-the-sky (POS) magnetic field orientation \citep{1979rpa..book.....R,1983Natur.304..243M,2001SSRv...99..243B,2007ASPC..365..242H,2009A&A...503..827X,2012SSRv..166..231R,2015A&ARv..24....4B,2016A&A...594A..10P,2019MNRAS.486.4813Z,2021ApJ...920....6G}. Further, combining synchrotron data across different frequencies has been proposed as a method to trace variations in the POS magnetic field along the line of sight (LOS). \cite{1966MNRAS.133...67B} introduced the Faraday Tomography (FT), suggesting that multi-layer magnetic field structures could be reconstructed via a proper Fourier transform of the synchrotron polarization data \citep{2005A&A...441.1217B}. \cite{2018ApJ...865...59L} noticed that synchrotron polarization structures tend to get elongated along the magnetic field, a consequence of anisotropy in magnetohydrodynamic (MHD) turbulence \citep{GS95,LV99,2012ApJ...747....5L,2016ApJ...818..178L}. This characteristic elongation serves as a tracer of the POS magnetic field orientation. Expanding upon this, the Synchrotron Polarization Gradient (SPG; \citealt{2018ApJ...865...59L}) technique has been developed. In the presence of Faraday Rotation, the polarized synchrotron emission is collected from a certain effective depth determined by the Faraday effect \citep{2016ApJ...818..178L,2018ApJ...865...59L,2019ApJ...887..258H}. SPG thus were introduced as a way to trace the POS magnetic field’s variation along the LOS by utilizing multi-frequency synchrotron polarization observations.

While the traditional methods focus on tracing the POS magnetic field orientation, \citet{2024arXiv240407806H} find that the 3D magnetic fields—which include the POS orientation, the magnetic field's inclination angle relative to the LOS, and magnetic field strength—are already encoded in synchrotron emission maps, because the anisotropy, or elongation along the magnetic field line, is inherently a 3D phenomenon. Therefore, the observed POS anisotropy, or the topology of synchrotron intensity structures, is affected by the projection effect, which is determined by the inclination angle, as well as by the magnetization level of the medium. 
Building upon these theoretical insights, \citet{2024arXiv240407806H} proposed using Convolutional Neural Networks (CNNs; \citealt{lecun1998gradient}) to extract the anisotropic features within synchrotron intensity maps and thereby enable the tracing of 3D magnetic fields. 

However, this CNN approach using synchrotron emission maps traces only the 3D magnetic fields averaged along the LOS. The magnetic fields' variations along the LOS are not determined. Thus, to fill the gap, in this work, we aim to develop a machine-learning approach to measure 3D magnetic fields' 3D spatial distribution. This is achieved by the combination with multi-wavelength synchrotron polarization observations which are subject to the Faraday decorrelation effect. This effect restricts the polarization information to a wavelength-dependent effective path length along the LOS \citep{2016ApJ...818..178L}. Consequently, the difference between synchrotron polarization at two wavelengths, i.e., the synchrotron polarization derivatives (SPDs), provides unique insights into the signal's spatial distribution along the LOS, enabling the reconstruction of the magnetic fields' 3D distribution.

Furthermore, we propose utilizing a Vision Transformer (ViT; \citealt{2017arXiv170603762V,2020arXiv201011929D}) model. The transformer architecture, widely recognized for its success in natural language processing through applications like the Chat Generative Pre-trained Transformer (ChatGPT; \citealt{Radford2018ImprovingLU,2020arXiv200514165B,radford2021learning}), has been adapted to ViT for 2D image processing. Compared with the CNN architecture, ViT's capability to handle complex data structures makes it particularly suitable for extracting and interpreting the anisotropic synchrotron structures that are correlated with 3D magnetic fields. Furthermore, our goal extends beyond algorithmic application; we seek an understanding of which features are indicative of magnetic field properties, why these features are significant, and the fundamental physical principles they represent.

This paper is organized as follows: \S~\ref{sec:theory} outlines the fundamental aspects of MHD turbulence anisotropy observed in synchrotron polarization and their correlation with 3D magnetic fields. It further describes our approach for capturing the 3D magnetic field variations along the LOS using multi-wavelength SPDs. \S~\ref{sec:data} provides an overview of the 3D MHD simulations and the synthetic synchrotron observations utilized in this study, including details of our ViT model. In \S~\ref{sec:result}, we present the results of the anisotropy analysis for SPDs and the 3D magnetic fields obtained from the ViT model. \S~\ref{sec:dis} delves into discussions on the potential of the ViT-SPDs methodology, as well as various physical problems that could benefit from insights into 3D magnetic fields. We conclude with a summary of our findings in \S~\ref{sec:con}.

\section{Theoretical consideration}
\label{sec:theory}
\subsection{Anisotropy in MHD turbulence: turbulent eddies are elongating along local magnetic fields}
Fluctuations induced by MHD turbulence were initially considered isotropic, disregarding the influence of magnetic fields \citep{1964SvA.....7..566I, 1965PhFl....8.1385K}. However, subsequent analytical \citep{GS95,LV99} and numerical studies \citep{2000ApJ...539..273C, 2001ApJ...554.1175M, 2003MNRAS.345..325C, 2010ApJ...720..742K, HXL21,2024MNRAS.527.3945H} as well as solar wind in situ measurements \citep{2016ApJ...816...15W, 2020FrASS...7...83M, 2021ApJ...915L...8D, 2023arXiv230512507Z} have provided the modern understanding of MHD turbulence with scale-dependent anisotropy. 

The revolutionary change in understanding this anisotropy is related to the "critical balance" condition, formulated by \cite{GS95} (hereafter GS95), which posits a balance between the turbulence cascading time and the Alfv\'en wave period:
\begin{equation}
(k_\bot \delta v_{l,\bot})^{-1}\approx(k_\parallel v_A)^{-1},
\end{equation}
where $v_A = B/\sqrt{4\pi\rho}$ is the Alfv\'en speed, with $B$ and $\rho$ being the magnetic field and gas density, respectively, while 
$k_\parallel$ and $k_\bot$ are the components of the wavevector parallel and perpendicular to the mean magnetic field.

However, this generally accepted picture requires an important correction. In LV99 picture of turbulent reconnection, magnetic turbulent eddies freely rotate perpendicular to the direction of the magnetic field percolating the eddy. This {\it local} magnetic field does not coincide with the mean magnetic field. Thus, all the measurements should be done in the local magnetic field reference frame:
\begin{equation}
    l_\bot^{-1} \delta v_{l,\bot}\approx l_\|^{-1} v_A, 
\end{equation}
 where $\delta v_{l,\bot}$ refers to the turbulent velocity at the scale $l_\bot$ measured in the local reference system, defined relative to the magnetic field intersecting an eddy. $l_\bot$ and $l_\parallel$ are the scales perpendicular and parallel to the magnetic field, respectively. The local system of reference was explicitly introduced in the pioneering study by \cite{2000ApJ...539..273C}. 

LV99 argues that turbulent reconnection of magnetic fields, occurring within an eddy turnover time, promotes magnetic line mixing perpendicular to their orientation, thus minimizing resistance to turbulent cascading. This process enables eddies to cascade efficiently perpendicular to the local magnetic field direction, which predominantly follows the Kolmogorov law in strong turbulence regime: $\delta v_{l,\bot} = (\frac{l_\bot}{L_{\rm inj}})^{1/3}\delta v_{\rm inj}M_A^{1/3}$, where $\delta v_{\rm inj}$ is the injection velocity at injection scale $L_{\rm inj}$ and $M_A=\delta v_{\rm inj}/v_A$ is the Alfv\'en Mach number. The strong turbulence regime in sub-Alfv\'enic turbulence ($M_A<$ 1) spans from the transitional scale $l_{\rm trans} = L_{\rm inj}M^2_A$ to smaller scales. Turbulence within the range from $L_{\rm inj}$ to $l_{\rm trans}$ is termed weak turbulence, which is wave-like and does not obey the "critical balance". The weak turbulence is also anisotropic, but the scaling relation is different from Eq.~\ref{eq.lv99} \citep{2001ApJ...562..279L,xu2019study}.

Adapting the "critical balance" condition in the local frame and integrating the Kolmogorov relationship in the strong turbulence regime, one can derive the the scale-dependent anisotropy scaling \citep{LV99}:
\begin{align}
\label{eq.lv99}
 l_\parallel= L_{\rm inj}(\frac{l_\bot}{L_{\rm inj}})^{\frac{2}{3}} M_A^{-4/3},~~~M_A\le 1,\\
\delta v_{l,\bot}= \delta v_{\rm inj}(\frac{l_\bot}{L_{\rm inj}})^{\frac{1}{3}}M_A^{1/3}, ~~~M_A\le 1.
\end{align}

Eq.~\ref{eq.lv99} reveals the anisotropic nature of turbulent eddies, with $l_\parallel \gg l_\bot$ for velocity fluctuation contours. Or equivalently, 
the anisotropy indicates more significant perpendicular than parallel velocity fluctuations at the same scales \citep{HXL21}. The relationships for density and magnetic field fluctuations can be derived from the linearized continuity and induction equations in Fourier space \citep{2003MNRAS.345..325C}:
\begin{align}
\label{eq.dBdrho}
        \omega \delta\rho_k = \rho_0 \pmb{k}\cdot \delta\pmb{v}_k,\\
        \omega\delta \pmb{B}_k = \pmb{k}\times(\pmb{B}_0\times \delta\pmb{v}_k),
\end{align}
where density $\rho$, magnetic field $\pmb{B}$, and velocity $v$ are described as a sum of their mean and fluctuating parts: $\rho=\rho_0+\delta\rho_l$, $\pmb{v}=\pmb{v}_0+\delta \pmb{v}_l$, and $\pmb{B}=\pmb{B}_0+\delta \pmb{B}_l$, where $\rho_0$ and $\pmb{B}_0$ denote the mean density and mean magnetic field strength, while the mean velocity field $\pmb{v}_0=0$.

\begin{figure*}
\includegraphics[width=0.99\linewidth]{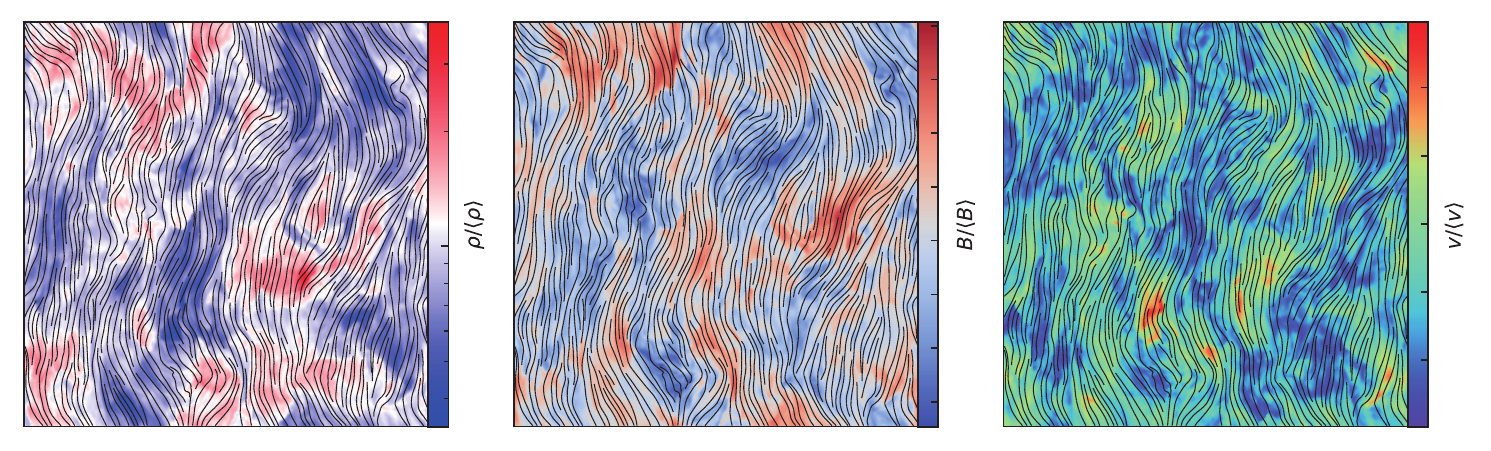}
        \caption{An numerical illustration of the anisotropy in the maps of gas density, magnetic field, and velocity slices. The black streamlines represent the magnetic field orientation. $\langle...\rangle$ denotes the mean value. Simulation with $M_A=0.79$ is used here. }
    \label{fig:rhoBvmap}
\end{figure*}

Considering the dispersion relation for Alfv\'enic turbulence is $\omega/k=v_A$ and the displacement vector $\pmb{\xi}$'s time derivative gives the velocity vector $\partial\pmb{\xi}/\partial t = \pmb{v}=v\hat{\pmb{\xi}}$, we obtain:
\begin{align}
\delta \rho_l&= \delta v_l\frac{\rho_0}{v_A}\mathcal{F}^{-1}(|\hat{\pmb{k}}\cdot\hat{\pmb{\xi}}|),\\
\delta B_l&= \delta v_l\frac{B_0}{v_A}\mathcal{F}^{-1}(|\hat{\pmb{B}}_0\times\hat{\pmb{\xi}}|), 
\end{align}
 where $\hat{\pmb{k}}$ and $\hat{\pmb{\xi}}$ represent the unit wavevector and displacement vector, respectively. $\mathcal{F}^{-1}$ denotes the inverse Fourier transform. The density and magnetic field fluctuations induced by turbulence are proportional to the velocity fluctuations and predominantly by their perpendicular components at given scales. One numerical example is given in Fig.~\ref{fig:rhoBvmap}. The structures/contours of density, magnetic field, and velocity are elongating along the local magnetic fields. {\it The elongation direction, thus, reveals the magnetic field orientation.}

For super-Alfv\'enic scenarios ($M_A \gg 1$), turbulence nears isotropy due to the diminished dynamic importance of magnetic fields. Turbulent energy cascades from larger to smaller scales, gradually reducing turbulent velocity until $M_A$ approximates unity at the transition scale $l_A$, derived as:
\begin{align}
\frac{1}{2}\rho(\frac{l_A}{L_{\rm inj}})^{2/3}\delta v_{\rm inj}^2&=\frac{1}{8\pi}B^2,\\
l_A&= L_{\rm inj}/M_A^3.
\end{align}

Below $l_A$ , magnetic field effects become significant, manifesting anisotropy as described by Eq.~\ref{eq.lv99}. However, above $l_A$, \cite{2024NatCo..15.1006H} noted a different type of anisotropy where weak magnetic fields, easily bent or influenced by turbulent flows, align with these flows. Consequently, the structures/contours of density, magnetic field, and velocity at large scales are still elongating along the local magnetic fields.  

\subsection{Anisotropy of MHD turbulence is imprinted in synchrotron emission and polarization}
The intrinsic relationship between synchrotron emission and the density of relativistic electrons and magnetic fields ensures that the properties of MHD turbulence are reflected in the observed synchrotron intensity and polarization structures. The observed intensity $I(\pmb{X},\lambda)$ and polarized emission $P(\pmb{X},\lambda^2)$ at wavelength $\lambda$ are expressed as \citep{1986rpa..book.....R,1970ranp.book.....P,2016ApJ...831...77L}:
\begin{align}
\label{eq.ip}
I(\pmb{X},\lambda) &= \int d\Omega \int_0^\infty dz I_i(\pmb{X},z,\lambda),\\
P(\pmb{X},\lambda^2) &= \int d\Omega \int_0^\infty dz(Q_i(\pmb{X},z,\lambda) + iU_i(\pmb{X},z,\lambda) ) e^{2i\lambda^2\phi(\pmb{X},z)},
\label{eq.ip2}
\end{align}
where $\pmb{X} = (x,y)$ represents spatial coordinates. 
The Faraday depth $\phi(\pmb{X},z)$ is given by \citep{2005A&A...441.1217B}:
\begin{equation}
\phi(\pmb{X},z) = 0.81 \int_0^z n_e(\pmb{X},z') B_\parallel(\pmb{X},z') dz' \quad \text{rad m}^{-2},
\end{equation}
where $n_e$ is the thermal electron number density in ${\rm cm^{-3}}$, $B_\parallel=B_z$ represents the LOS  magnetic field component in $\mu$G, and the path length $z$ is measured in parsecs. The intrinsic synchrotron emission $I_i(\pmb{X},z,\lambda)$ and intrinsic polarization $P_i(\pmb{X},z,\lambda)=Q_i+iU_i$, described by the Stokes parameters $Q_i$ and $U_i$, are given by:
\begin{align}
\label{eq.iqu_i}
I_i(\pmb{X},z,\lambda) &\propto \lambda^{\frac{p-1}{2}} n_{e,r} (B_x^2+B_y^2)^{\frac{p-3}{4}} (B_x^2+B_y^2),\\
\label{2}
Q_i(\pmb{X},z,\lambda) &\propto \lambda^{\frac{p-1}{2}} n_{e,r} (B_x^2+B_y^2)^{\frac{p-3}{4}} (B_x^2-B_y^2),\\
\label{3}
U_i(\pmb{X},z,\lambda) &\propto \lambda^{\frac{p-1}{2}} n_{e,r} (B_x^2+B_y^2)^{\frac{p-3}{4}} (2B_xB_y),
\end{align}
where $n_{e,r}(\pmb{X},z)$ indicates the relativistic electron number density, and $p$ denotes the spectral index of the electron energy distribution $E$:
\begin{equation}
\label{eq.NE}
N(E) dE = N_0 E^{-p}dE,
\end{equation}
with $N(E)$ representing the electron number density per unit energy interval $dE$. The pre-factor $N_0$ is derived by integrating Eq.~\ref{eq.NE} to obtain the total electron number density and assume a lower limit for $E$.


Moreover, as indicated in Eq.~\ref{eq.dBdrho}, when describing the density and magnetic field as a sum of their mean and fluctuating components, their fluctuations are predominantly perpendicular to the magnetic field, due to the anisotropy of MHD turbulence (see Eq.~\ref{eq.lv99}). The expressions in Eq.~\ref{eq.iqu_i} suggest that fluctuations in synchrotron emission intensity and polarization are determined by these in the magnetic field and density. 
The dependence of the synchrotron properties on the spectral index $p$ was studied by \cite{2012ApJ...747....5L}. There, synchrotron fluctuations for an arbitrary index are analytically shown to be anisotropic reflecting the statistics of MHD turbulence. Other constant factors at a given $p$ are not explicitly detailed in Eq.~\ref{eq.iqu_i}, as they do not alter the characteristics of these fluctuations. Consequently, the fluctuations in $I_i(\pmb{X},z,\lambda)$ and $P_i(\pmb{X},z,\lambda)$ exhibit pronounced anisotropy, showing more significant fluctuations perpendicular to the magnetic field. This anisotropy implies that the contours of synchrotron intensity and polarization—essentially the structures of these emissions—elongate along the magnetic field lines, thereby indicating the orientation of the magnetic field.

\begin{figure*}
	\includegraphics[width=1.0\linewidth]{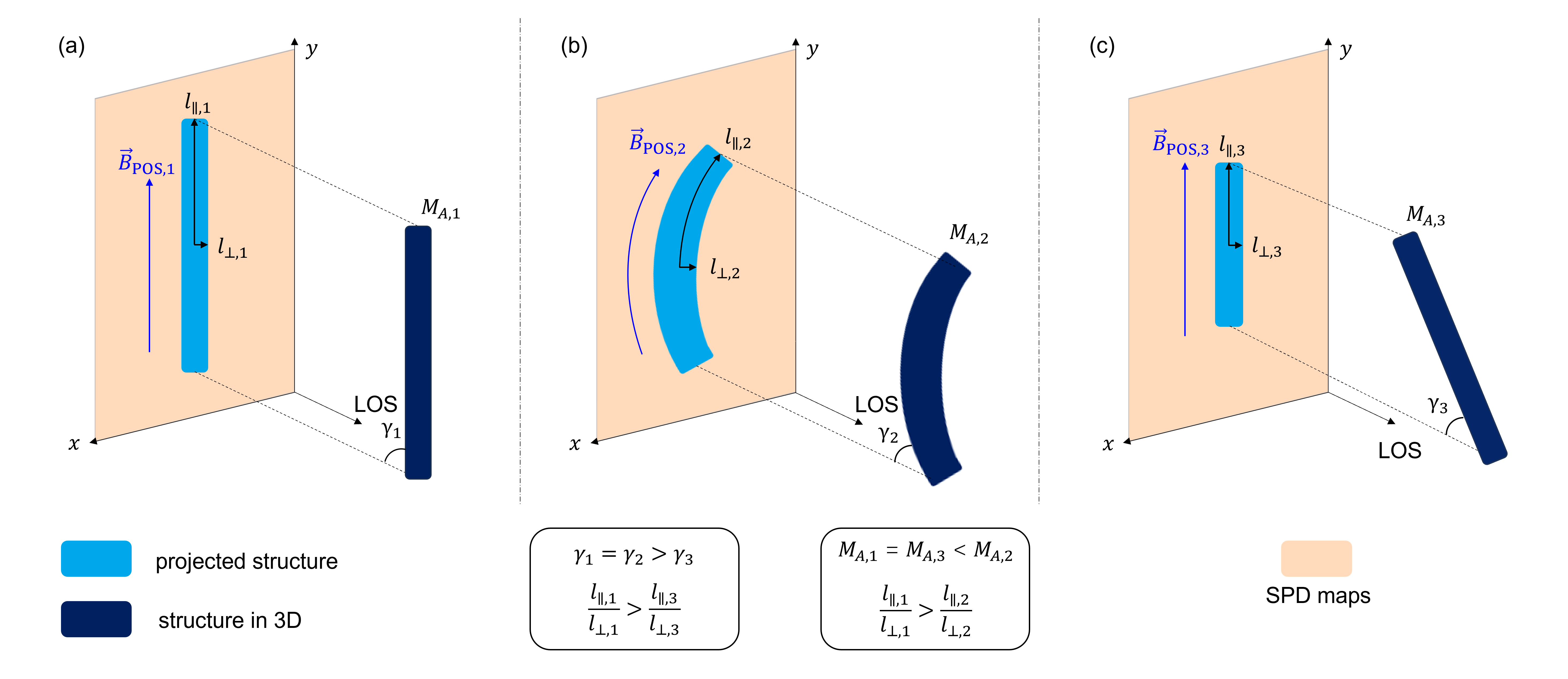}
    \caption{An Illustration of how the observed structures in synchrotron polarization's derivative maps are regulated by the Alfv\'en Mach number $M_A$ and inclination angle $\gamma$. Within all three panels, these intensity structures elongate along the POS magnetic field orientation where $l_\parallel>l_\bot$. Structures 1 and 2, depicted in panels (a) and (b), are projected onto the POS with identical inclination angles $\gamma_1=\gamma_2$, yet exhibit different magnetizations with $M_{A,1}^{-1}>M_{A,2}^{-1}$. Notably, the anisotropy observed, represented as $l_\parallel/l_\bot$, in the weakly magnetized Structure 2 is less pronounced than in Structure 1. Structure 2 is less straightened because the weak magnetic field has more fluctuations. Comparatively, Structures 1 and 3—showcased in panels (a) and (c)—possess equivalent magnetizations $M_{A,1}^{-1}=M_{A,3}^{-1}$, but different inclination angles with $\gamma_1>\gamma_3$. The observed anisotropy decreases with smaller  $\gamma$, though it is crucial to note that the straightness of Structure 3 remains unaffected by this projection. Modfied from \citet{2024MNRAS.52711240H}.}
    \label{fig:anisotropy}
\end{figure*}


\subsection{Projection effect: retrieving 3D magnetic field information}
\label{sec:SPD}
\subsubsection{Information on $\gamma$}
The anisotropy in intrinsic synchrotron emission and intrinsic polarization (see Eq.~\ref{eq.iqu_i}) suggests that the structures of either synchrotron emission or synchrotron polarization are elongating along the 3D magnetic fields in 3D space. However, in addition to the effect of Faraday rotation, the observed polarization is subject to the projection along the LOS. The observed synchrotron structures are therefore elongating along the projected magnetic fields on the POS. 

\textbf{Anisotropy degree:} The projection effect alters the observed anisotropy degree, defined as $l_\parallel/l_\bot$. Fig.~\ref{fig:anisotropy} presents a cartoon illustration of how the 3D magnetic field's inclination angle $\gamma$ relative to the LOS changes the observed anisotropy. Structures 1 and 3—showcased in panels (a) and (c)—possess equivalent magnetizations, but different inclination angles with $\gamma_1>\gamma_3$. The observed anisotropy of Structure 1 is therefore smaller than that of Structure 3. Therefore, the elongation direction $l_\parallel$ gives the POS magnetic field orientation, while the observed anisotropy degree contains the information on the magnetic field's inclination angle. This also modifies the observed anisotropy as discussed in \cite{2012ApJ...747....5L}.

\textbf{Curvature:}  On the other hand, in the presence of fluctuations, magnetic field lines are not straight but exhibit curvature. This curvature is naturally inherited by the elongation of synchrotron structures along the curved magnetic fields in 3D space. However, the curvature of both the magnetic field lines and the elongation of these structures is also subject to the projection effect. Consequently, the morphological curvature in the observed synchrotron structures provides additional insights into the magnetic field's inclination angle.

\subsubsection{Information on $M_A$}
\textbf{Anisotropy degree:} The observed degree of anisotropy is influenced not only by projection effects but also by the medium's magnetization, defined as $M_A^{-1}$. According to Eq.~\ref{eq.lv99}, the degree of anisotropy in synchrotron structures is more pronounced in environments with strong magnetization. Conversely, a weak magnetization leads to a lower degree of anisotropy as the magnetic field's back-reaction on turbulence becomes less dynamically significant.

\textbf{Curvature:} Additionally, as illustrated in Fig.~\ref{fig:anisotropy}, the magnetization of the medium is also reflected in the curvature of the magnetic field lines. In a weakly magnetized medium, significant magnetic field fluctuations allow for easier bending of the magnetic field lines, resulting in more pronounced curvature in both the field lines and the associated synchrotron structures. Within a domain of strong magnetization, magnetic field lines exhibit minimal deviation resulting in predominantly straightened topology.

In summary, the observed anisotropy in synchrotron emission or polarization (also its derivative) has three important implications: 
\begin{enumerate}
    \item Synchrotron structures are predominantly elongated along the local magnetic field lines, making the elongation an indirect probe of the POS magnetic field orientation.
    \item The observed degree of anisotropy is closely associated with the magnetization level and the magnetic field’s inclination angle. 
    \item The observed morphological curvature of synchrotron structures is sensitive to both magnetization and inclination angle. 
\end{enumerate}
Consequently, by analyzing the elongation direction, the degree of anisotropy, and the morphological curvature of the observed synchrotron structures, we can retrieve the 3D magnetic field information, including the POS orientation, the magnetization level, and the inclination angle.
 
\begin{figure*}
 \centering
\includegraphics[width=0.85\linewidth]{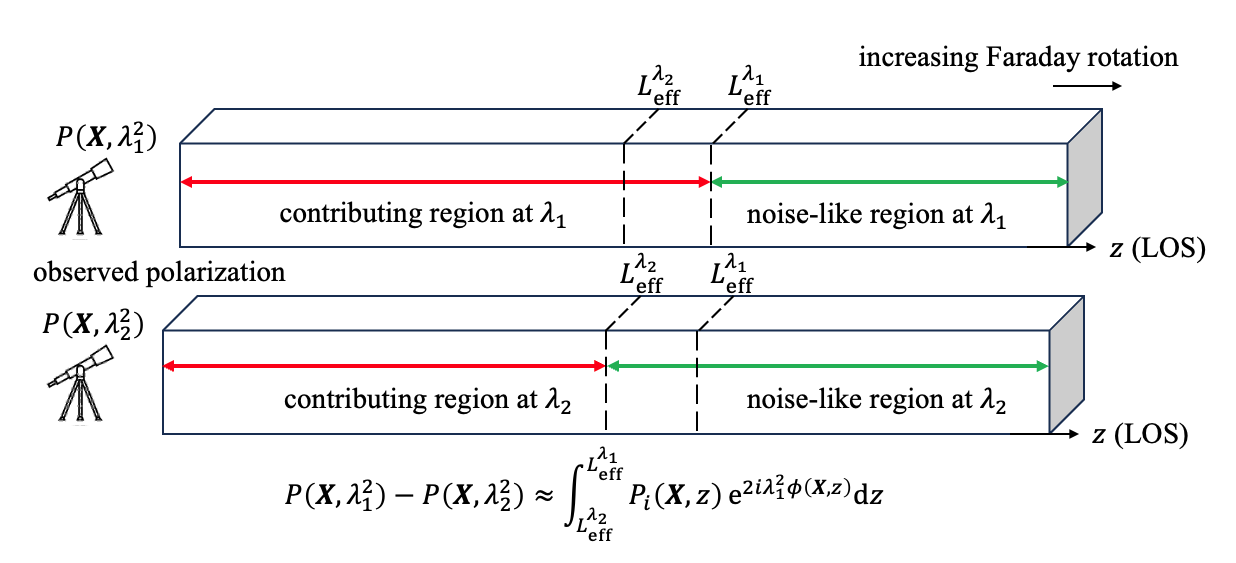}
        \caption{Cartoon illustration of the information provided by the difference between polarized synchrotron intensity maps $P(\pmb{X},\lambda^2)$. The tube represents a synchrotron emitting region with the presence of Faraday rotation but observed at two wavelengths $\lambda_1$ and $\lambda_2$. The effective length scales of $P(\pmb{X},\lambda^2_1)$ and $P(\pmb{X},\lambda^2_2)$ are $L^{\lambda_1}_{\rm eff}$ and $L^{\lambda_2}_{\rm eff}$, respectively, along the LOS. Beyond $L^{\lambda_1}_{\rm eff}$ and $L^{\lambda_2}_{\rm eff}$, the noise-like regions are dominantly contaminated by Faraday rotation. The difference in $P(\pmb{X},\lambda^2_1)$ and $P(\pmb{X},\lambda^2_2)$ gives the polarized intensity between $L^{\lambda_1}_{\rm eff}$ and $L^{\lambda_2}_{\rm eff}$. }
    \label{fig:ill_p}
\end{figure*}

 \begin{figure*}
  \centering
\includegraphics[width=1.0\linewidth]{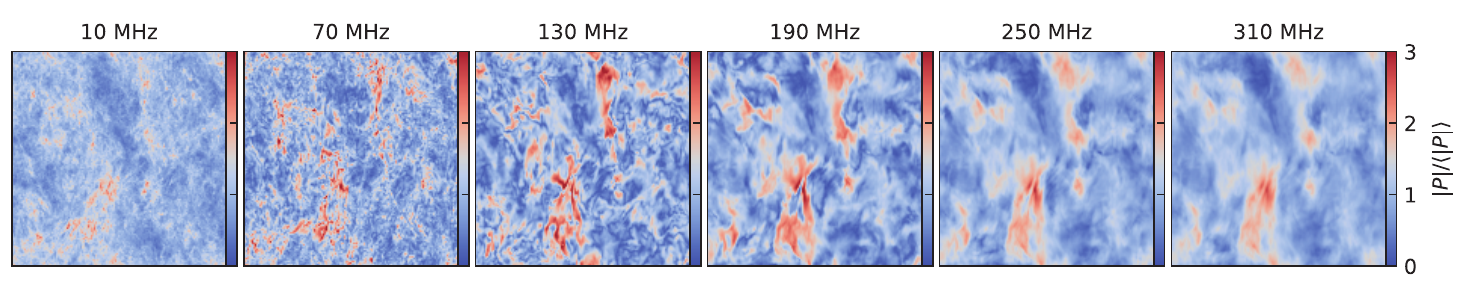}
        \caption{An numerical illustration of the changes in polarized synchrotron intensity map $|P(\pmb{X},\lambda^2)|$ regarding different frequencies. The changes are induced by Faraday rotation whose significance increases at low frequencies. Simulation with  $M_A=1.21$ and $\gamma=90^\circ$ is used here. }
    \label{fig:num_p}
\end{figure*}
\subsection{Effect of Faraday rotation: obtaining 3D spatial information from synchrotron polarization derivative}
The observed polarization, encoded with the anisotropy of MHD turbulence, is affected by the Faraday rotation, which depends on thermal electron number density, LOS magnetic field strength, and the integration length. The Faraday rotation becomes increasingly significant along the LOS. This rotation results in Faraday depolarization, reducing the observed polarization fraction, and Faraday decorrelation, diminishing the correlation of polarization fluctuations. {\it Beyond the effective path length $L_{\rm eff}$, the Faraday decorrelation effect is substantial such that the fluctuations of polarized synchrotron emission become uncorrelated, contributing only noise-like signals. These signals lose the statistical properties of MHD turbulence.} The condition for determining $L_{\rm eff}$ is expressed as \citep{2016ApJ...818..178L}:
\begin{equation}
\label{eq.Leff}
\phi(\pmb{X},z) \lambda^2 = 0.81 \lambda^2 \int_0^{L_{\rm eff}} n_e(\pmb{X},z) B_\parallel(\pmb{X},z) dz \approx \pm1 \text{ rad},
\end{equation}
indicating that $L_{\rm eff}$ is wavelength-dependent \footnote{ It should be noted that the one-radian criterion used in Eq.~\ref{eq.Leff} is applied for simplicity, with the full expression provided in \cite{2016ApJ...818..178L}. While the exact value of Eq.~\ref{eq.Leff} may vary, but the underlying physical meaning remains valid—specifically, that the polarization signal becomes de-correlated beyond an effective length scale, $L_{\rm eff}$.}. Based on this criterion, weak and strong Faraday decorrelation regimes are distinguished by $L_{\rm eff}/L < 1$ and $L_{\rm eff}/L > 1$, respectively. Only the region where $L_{\rm eff}/L < 1$ contributes useful information to the observed polarization.
 
Fig.~\ref{fig:ill_p} illustrates a synchrotron emitting region observed at two different wavelengths, 
$\lambda_1$ and $\lambda_2$. The effective lengths of polarization, $P_1$ and $P_2$, correspond to $L^{\lambda_1}_{\rm eff}$ and $L^{\lambda_2}_{\rm eff}$, respectively. Regions beyond $L^{\lambda_1}_{\rm eff}$ and $L^{\lambda_2}_{\rm eff}$  are predominantly contaminated by strong Faraday rotation, effectively introducing only noise. Therefore, $P_1$ and $P_2$ collect polarization signals only up to  $L^{\lambda_1}_{\rm eff}$ and $L^{\lambda_2}_{\rm eff}$, respectively. A numerical example of the polarized synchrotron emission map at different frequencies is given in Fig.~\ref{fig:num_p}. At a low frequency of 10 MHz, the map exhibits significant Faraday rotation, resulting in a predominantly noise-like observed polarization due to the decorrelation of the fluctuations. In contrast, the high-frequency map at 310 MHz experiences minimal Faraday rotation, allowing the observed polarization map to sample a longer path length along the LOS. Consequently, the structures in this high-frequency map appear more coherent and correlated.

The effect of Faraday decorrelation allows for the use of multi-wavelength observations to sample only the emitting region between $L^{\lambda_1}_{\rm eff}$ and $L^{\lambda_2}_{\rm eff}$, thus obtaining 3D spatial information about the polarization. This is facilitated by calculating the synchrotron polarization derivative (SPD). When the difference, $\Delta\lambda^2 = \lambda_1^2 - \lambda_2^2$, between $\lambda_1$ and $\lambda_2$ is not small, the zero-order approximation yields \footnote{ Note here the integration considers only the contribution from correlated polarization up to $L_{\rm eff}$. Beyond $L_{\rm eff}$, the uncorrelated polarization, contributing only noise-like signals, is ignored.}:
\begin{equation}
\begin{aligned}
\label{eq.dp0th}
\frac{d P(\pmb{X},\lambda^2)}{d \lambda^2} &\approx \frac{P(\pmb{X},\lambda_1^2) - P(\pmb{X},\lambda_2^2)}{\lambda_1^2 - \lambda_2^2} \\
&= \frac{\int d\Omega\int_{L^{\lambda_2}_{\rm eff}}^{L^{\lambda_1}_{\rm eff}}P_i(\pmb{X},z)e^{2i\lambda^2_1\phi(\pmb{X},z)} dz}{\Delta\lambda^2 },
\end{aligned}
\end{equation}
thereby quantifying the variations in polarization due to differences in wavelength and elucidating the spatial distribution of polarization across different depths. Alternatively, when $\lambda_1$ closely approaches $\lambda_2$, the derivative is given by:
\begin{equation}
\begin{aligned}
\label{eq.dp1st}
\frac{d P(\pmb{X},\lambda^2)}{d \lambda^2} &= \lim_{\lambda_1 \to \lambda_2} \frac{P(\pmb{X},\lambda_1^2) - P(\pmb{X},\lambda_2^2)}{\lambda_1^2 - \lambda_2^2} \\
&= 2i\int_0^{L^{\lambda_1}_{\rm eff}} P_i(\pmb{X},z)\phi(\pmb{X},z)e^{2i\lambda_1^2\phi(\pmb{X},z)}dz,
\end{aligned}
\end{equation}
where $P_i(\pmb{X},z)$ is weighted by $\phi(\pmb{X},z)$, enhancing the emphasis on contributions from the Faraday rotation measure. For broader observational application and simplicity, we focus on the synchrotron polarization derivative as defined in Eq.~\ref{eq.dp0th}.


\section{Methodology}
\label{sec:data}

\subsection{MHD simulations and synthetic synchrotron observations}
 The MHD numerical simulations used in this study were conducted using the ZEUS-MP/3D code \citep{2006ApJS..165..188H}. Considering periodic boundary conditions, we generate an isothermal simulation of MHD turbulence, solving the ideal MHD equations within an Eulerian framework. Kinetic energy was solenoidally injected at a wavenumber of 2 to simulate a Kolmogorov-like power spectrum. The computational domain was discretized into a grid of $792^3$ cells, with numerical turbulence dissipation manifesting at scales ranging from approximately 10 to 20 cells.
  
Initial conditions for the simulations include a uniform density field with a uniform magnetic field oriented along the $y$-axis. Subsequently, the simulation cubes are rotated to align the mean magnetic field inclination relative to the LOS, or the $z$-axis, at angles of $90^\circ$, $60^\circ$, and $30^\circ$. Scale-free turbulence simulations are characterized by the sonic Mach number, $M_s = \delta v_{\rm inj}/c_s$, and the Alfv\'enic Mach number, $M_A = \delta v_{\rm inj}/v_A$. Here $c_s$ is the isothermal sound speed. We generate a range of $M_A$ and $M_s$ values to accommodate various physical scenarios, as listed in Tab.~\ref{tab:sim}. More details of the simulation setup can be found in \cite{2024MNRAS.52711240H,2024arXiv240407806H}.

\begin{table}
	\centering
 \begin{tabular}{ | c | c | c | c | c |}
		\hline
		Run & $M_s$ & $M_A$ & range of $M_{A,\lambda}^{\rm sub}$  & range of $M_{s,\lambda}^{\rm sub}$\\ \hline \hline
		Z0 & 0.66 & 0.26 & 0.02 - 0.30 & 0.37 - 0.91 \\ 
		Z1 & 0.62 & 0.50 & 0.06 - 0.62 & 0.37 - 0.89 \\
		Z2 & 0.61 & 0.79 & 0.11 - 1.32 & 0.38 - 0.82 \\ 
		Z3 & 0.59 & 1.02 & 0.15 - 1.99 & 0.37 - 0.80 \\ 
		Z4 & 0.58 & 1.21 & 0.21 - 2.19 & 0.38 - 0.82 \\\hline
	\end{tabular}
	\caption{\label{tab:sim}$M_s$ and $M_A$ are the sonic Mach number and the Alfv\'enic Mach number calculated from the global injection velocity, respectively. $M_{A,\lambda}^{\rm sub}$ and $M_{s,\lambda}^{\rm sub}$ are determined using the local velocity dispersion calculated along each LOS in a $22\times22$ cells sub-field. 
 }
\end{table}

\begin{figure*}
\centering
\includegraphics[width=1.0\linewidth]{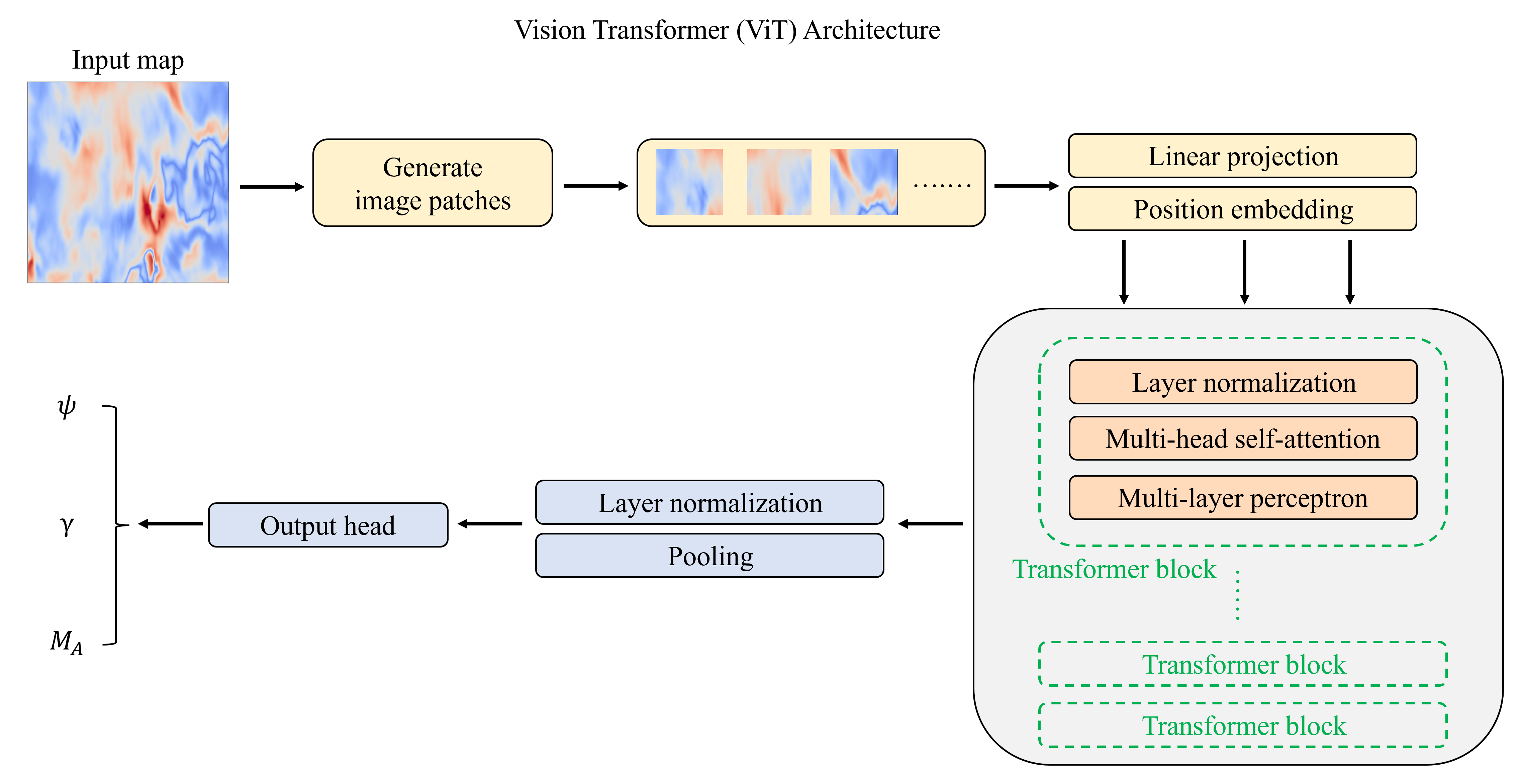}
        \caption{Architecture of the ViT-model. The input image is a $22\times22$-cells map cropped from the synchrotron polarization derivative map. The network outputs the magnetic field's POS orientation angle $\psi$, inclination angle $\gamma$, and the Alfv\'en Mach number $M_A$.}
    \label{fig:ViT}
\end{figure*}

To generate synthetic synchrotron observations from our simulations, we use the density field
$\rho(\pmb{x})$ and the magnetic field $\pmb{B}(\pmb{x})$. The calculations for synchrotron intensity 
$I(\pmb{X})$ and polarization $P(\pmb{X},\lambda^2)$ are based on Eqs.~\ref{eq.ip} and \ref{eq.ip2}. Given the relative insensitivity of synchrotron emission to variations in the electron energy distribution's spectral index, as noted by \cite{2012ApJ...747....5L, 2019ApJ...886...63Z}, we assume a homogeneous and isotropic electron energy distribution with a spectral index $p=3$. 

\subsection{Vision Transformer (ViT)}
\subsubsection{ViT architecture}
In this work, we design a Vision Transformer (ViT; \citealt{2017arXiv170603762V,2020arXiv201011929D}) model to process synchrotron maps, extract morphological features, and output the magnetic field's POS orientation angle $\psi$, inclination angle $\gamma$, and the Alfv\'en Mach number $M_A$, by using the transformer architecture (see Fig.~\ref{fig:ViT}). It starts with an input layer that accepts maps of synchrotron polarization derivative. The ViT can be separated into three stages:

\textbf{Feature Extraction Stage:} The input layer accepts maps of synchrotron polarization derivatives. The map is divided into smaller, non-overlapping patches, which contain information on local features. The patches are then flattened and linearly projected into a higher-dimensional space, creating patch embeddings \citep{2017arXiv170603762V}. Positional encodings are added to these embeddings to retain spatial information about the original map structure. 

\textbf{Correlation Learning Stage:} The sequence of patch embeddings is processed through eight transformer blocks. Each block comprises a normalization step \citep{2016arXiv160706450L} followed by a multi-head self-attention mechanism \citep{2017arXiv170603762V}, allowing the model to focus on different parts of the input simultaneously. After the attention mechanism, a residual connection is added, and the output is passed through a multi-layer perceptron (MLP; \citealt{Goodfellow-et-al-2016}) that includes two dense layers with a non-linear activation function and dropout for regularization. Another residual connection \citep{he2015deepresiduallearningimage} follows this MLP, ensuring effective model training. During this stage, the model learns the correlation of local features with the magnetic field.

\textbf{Output Stage:} The final output of the transformer layers undergoes layer normalization and global average pooling \citep{2013arXiv1312.4400L} to condense the feature information into a single vector. This vector is then passed through a dense layer, known as the classification head, which outputs $\psi$, $\gamma$, and $M_A$.

Compared with the earlier CNN models \citep{2024MNRAS.52711240H,2024arXiv240407806H}, which rely on convolutional layers to capture local features and hierarchical representations, the ViT model uses self-attention mechanisms to integrate information across the entire input map. That means every patch of the map can directly interact with and learn from every other part, instead of relying on only single local information. ViT, thus,
can provide a more holistic understanding of the data and a more accurate estimation.

\subsubsection{ViT training strategies}
The trainable parameters of the ViT are optimized using a typical deep learning approach, where the mean absolute error of the 3D magnetic field prediction serves as the training loss function for backpropagation \citep{rumelhart1986learning}. Additionally, we implement two strategies to enhance the diversity and randomness of the training dataset \citep{doi:10.1198/10618600152418584}, which are essential for improving the ViT’s generalization across varying physical conditions.

\textbf{Random Cropping:} One effective technique is random cropping \citep{2018arXiv181109030T}, which involves extracting smaller patches of size $22\times22$ cells from the input images. This method not only enlarges the dataset but also introduces varied perspectives, thereby enhancing the model's exposure to different features within the maps of synchrotron polarization derivative. The choice of $22\times22$ cells is deliberate to mitigate the effects of numerical turbulence dissipation and maintain high-resolution prediction. As studied by \citet{2024arXiv240407806H}, different sizes do not impact deep learning models' accuracy after sufficient training, but only the output magnetic field map's resolution.
 \begin{figure*}
 \centering
\includegraphics[width=0.99\linewidth]{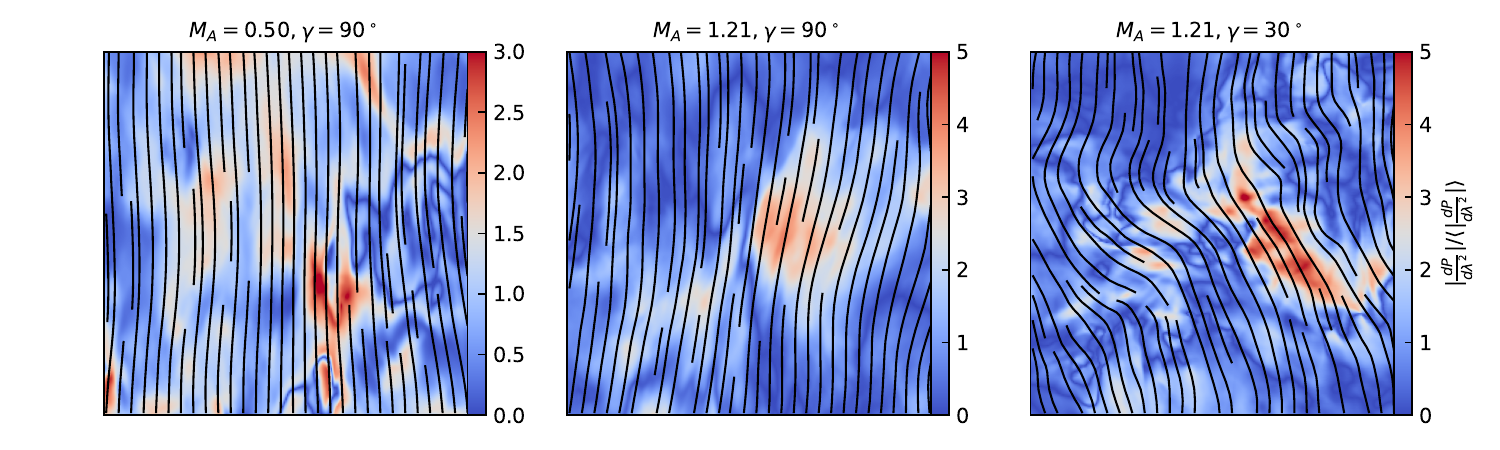}
        \caption{A numerical illustration of the anisotropy in the maps of synchrotron polarization's derivative $|\frac{d P(\pmb{X},\lambda^2)}{d \lambda^2}|$. The derivative is calculated between 280 MHz and 310 MHz polarization maps. The black streamlines represent the POS magnetic field orientation. Left panel: $M_A=0.50$, $\gamma=90^\circ$. 
        Middle panel: $M_A=1.21$, $\gamma=90^\circ$. Right panel: $M_A=1.21$, $\gamma=30^\circ$. }
    \label{fig:2Dmap}
\end{figure*}

\begin{figure}
 \centering
\includegraphics[width=1.0\linewidth]{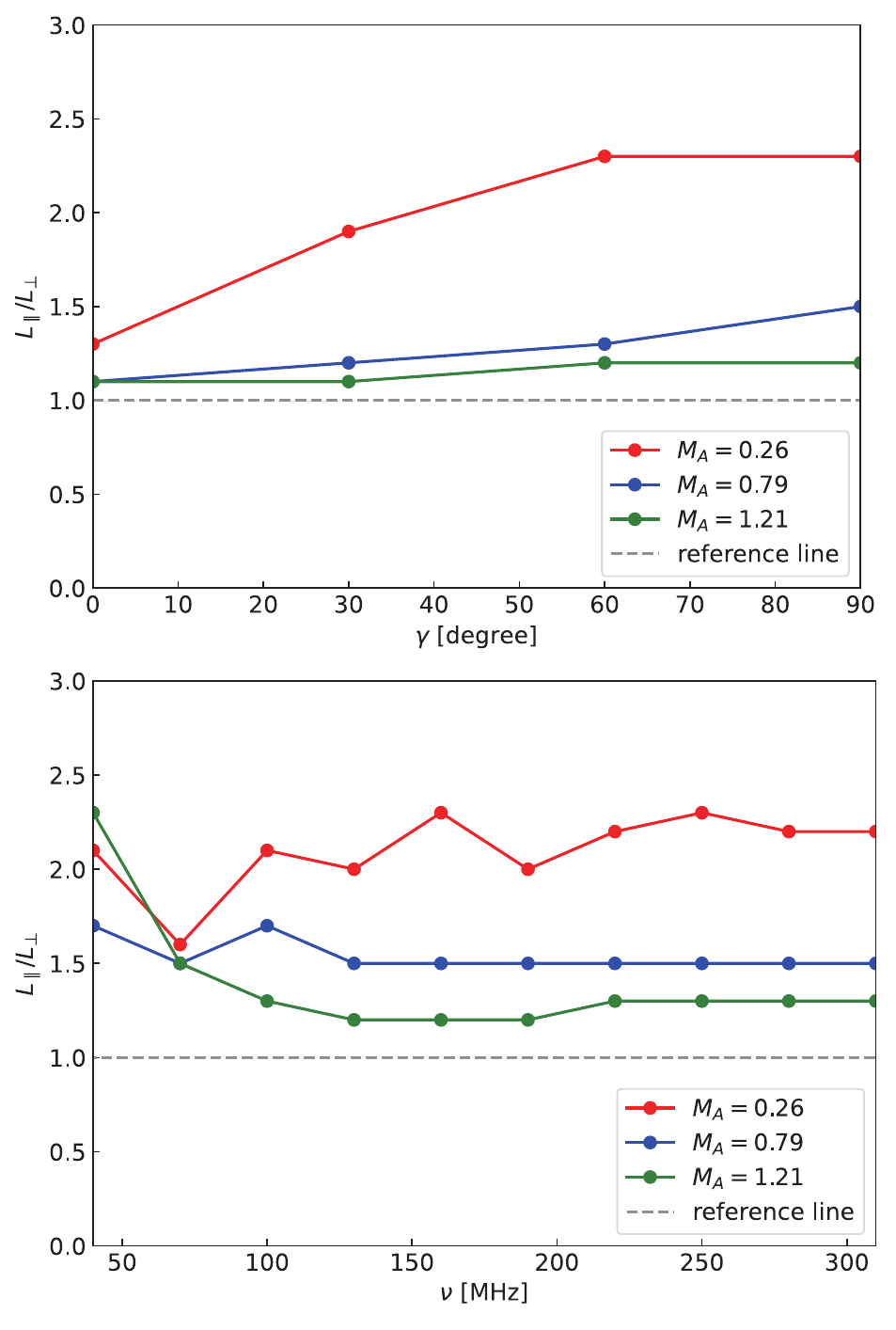}
        \caption{The correlation of $|\frac{d P(\pmb{X},\lambda^2)}{d \lambda^2}|$'s anisotropy ratio $L_\parallel/L_\bot$ with the magnetic field's inclination angle $\gamma$, Alfv\'en Mach number $M_A$, and the observation frequency $\nu$. Upper panel: the derivative is calculated between 280 MHz and 310 MHz polarization maps. Lower panel: these derivatives are calculated across a frequency range from 10 MHz to 310 MHz in increments of 30 MHz. $\gamma=90$ degrees. }
    \label{fig:anisotropy_num}
\end{figure}

\textbf{Random Rotation:} Furthermore, images lack rotational invariance from the computational model’s perspective. Since each image cell corresponds to an element in a matrix, rotating an image changes the matrix’s element arrangement, presenting the image as novel data to the model \citep{10.1145/1273496.1273556}. This property is utilized in two ways: first, by randomly rotating the $22\times22$-cell patches to augment the training dataset further, and second, by using the original, unrotated datasets for validation, thereby generating a prediction test scenario.

Approximately 0.2 million sub-fields are fed into the ViT model for each training iteration. We conducted 100 training iterations, continuing until the loss function indicated saturation, suggesting that the model parameters had converged effectively. 

\subsection{Training images}
Our training input is a SPD map, $|\frac{d P(\pmb{X},\lambda^2)}{d \lambda^2}|$. The $792\times792$-cells SPD map is randomly segmented into $22\times22$-cell subfields for input into the ViT model. For each subfield, we also generate corresponding projected maps of $\psi^{\rm sub}_{\lambda_1}$, $\gamma^{\rm sub}_{\lambda_1}$, $M_{A,\lambda_1}^{\rm sub}$, and $M_{s,\lambda_1}^{\rm sub}$ as per the following:
\begin{align}
\psi^{\rm sub}_{\lambda_1}(\pmb{X})&=\arctan(\frac{\int_{L^{\lambda_2}_{\rm eff}}^{L^{\lambda_1}_{\rm eff}} B_y(\pmb{X},z)dz}{\int_{L^{\lambda_2}_{\rm eff}}^{L^{\lambda_1}_{\rm eff}}B_x(\pmb{X},z)dz}),\\
\gamma^{\rm sub}_{\lambda_1}(\pmb{X})&=\arccos(\frac{\int_{L^{\lambda_2}_{\rm eff}}^{L^{\lambda_1}_{\rm eff}} B_z(\pmb{X},z)dz}{\int_{L^{\lambda_2}_{\rm eff}}^{L^{\lambda_1}_{\rm eff}} B(\pmb{X},z)dz}),\\
M_{A,\lambda_1}^{\rm sub}(\pmb{X})&=\frac{v_{\lambda_1}^{\rm sub}\sqrt{4\pi\langle\rho\rangle^{\rm sub}_{\lambda_1}}}{\langle B\rangle^{\rm sub}_{\lambda_1}}, \\
M_{s,\lambda_1}^{\rm sub}(\pmb{X})&=\frac{v_{\lambda_1}^{\rm sub}}{c_s},
\end{align}
where $B=\sqrt{B_x^2+B_y^2+B_z^2}$ is the total magnetic field strength, and $B_x$, $B_y$, and $B_z$ are its $x$, $y$, and $z$ components. $L^{\lambda_1}_{\rm eff}$ and $L^{\lambda_2}_{\rm eff}$ are determined by the conditions given in Eq.~\ref{eq.Leff}.
$\lambda_1$ and $\lambda_2$ are chosen so that the difference in their corresponding frequencies is 30 MHz. $\langle\rho\rangle^{\rm sub}_{\lambda_1}$ and $\langle B\rangle^{\rm sub}_{\lambda_1}$ are the mass density and magnetic field strength averaged within the range of $L^{\lambda_1}_{\rm eff}$ and $L^{\lambda_2}_{\rm eff}$ for every sub-field. $M_{A,\lambda_1}^{\rm sub}$ and $M_{s,\lambda_1}^{\rm sub}$ are defined using the local velocity dispersion for each sub-field between $L^{\lambda_1}_{\rm eff}$ and $L^{\lambda_2}_{\rm eff}$ along the LOS (i.e., $v_{\lambda_1}^{\rm sub}$), rather than the global turbulent injection velocity $v_{\rm inj}$ used to characterize the full simulation. The ranges of $M_{A,\lambda_1}^{\rm sub}$ and $M_{s,\lambda_1}^{\rm sub}$ averaged over the subfield in each simulation are listed in Tab.~\ref{tab:sim}, while the range of local inclination angle $\gamma^{\rm sub}_{\lambda_1}$ spans from 0 to 90$^\circ$. These values of $M_{A,\lambda_1}^{\rm sub}$, $M_{s,\lambda_1}^{\rm sub}$, and $\gamma^{\rm sub}_{\lambda_1}$ cover typical physical conditions of diffuse medium.

\begin{figure*}
\centering
\includegraphics[width=1.0\linewidth]{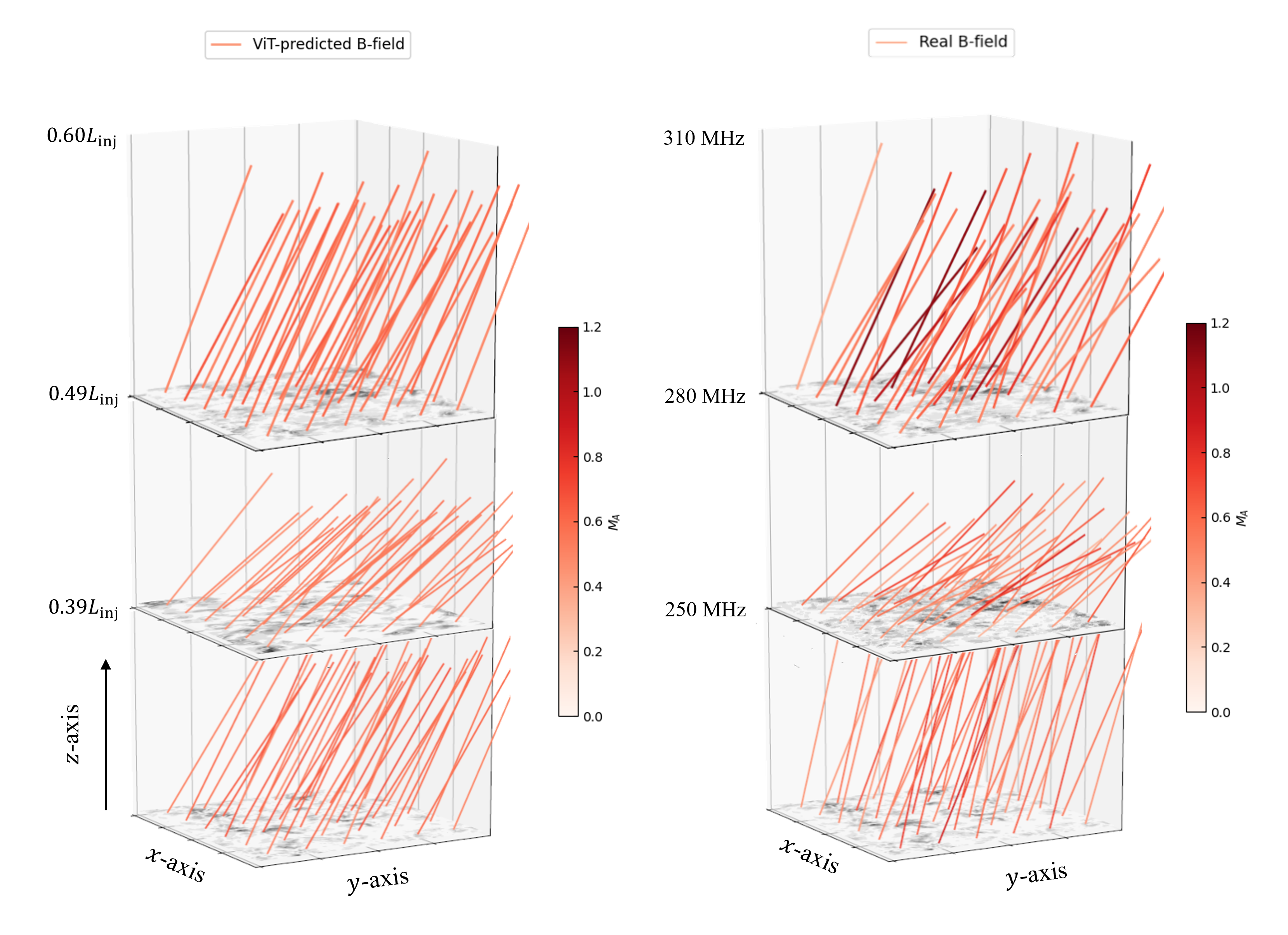}
            \caption{A comparison of the ViT-predicted 3D magnetic fields (left) and actual 3D magnetic field (right) obtained from simulation. Each segment of the magnetic field is constructed from the POS position angle and the inclination angle. The simulation parameters include a global $M_A=0.79$ and $M_s=0.61$. Local sub-field $M_{A,\lambda_1}^{\rm sub}$ are derived using small-scale local velocity dispersions specific to sub-fields between $L^{\lambda_1}_{\rm eff}$ and $L^{\lambda_2}_{\rm eff}$ along the LOS. $M_{A,\lambda_1}^{\rm sub}$ is thus smaller than the global $M_A$. SPD maps at frequencies of 310, 280, and 250 MHz are used. Each visualized magnetic field segment is averaged over 132×132 pixels. The SPD maps are positioned on the POS, corresponding to the $x-y$ plane.}
    \label{fig:3D_sub13}
\end{figure*}

\begin{figure*}
\centering
\includegraphics[width=1.0\linewidth]{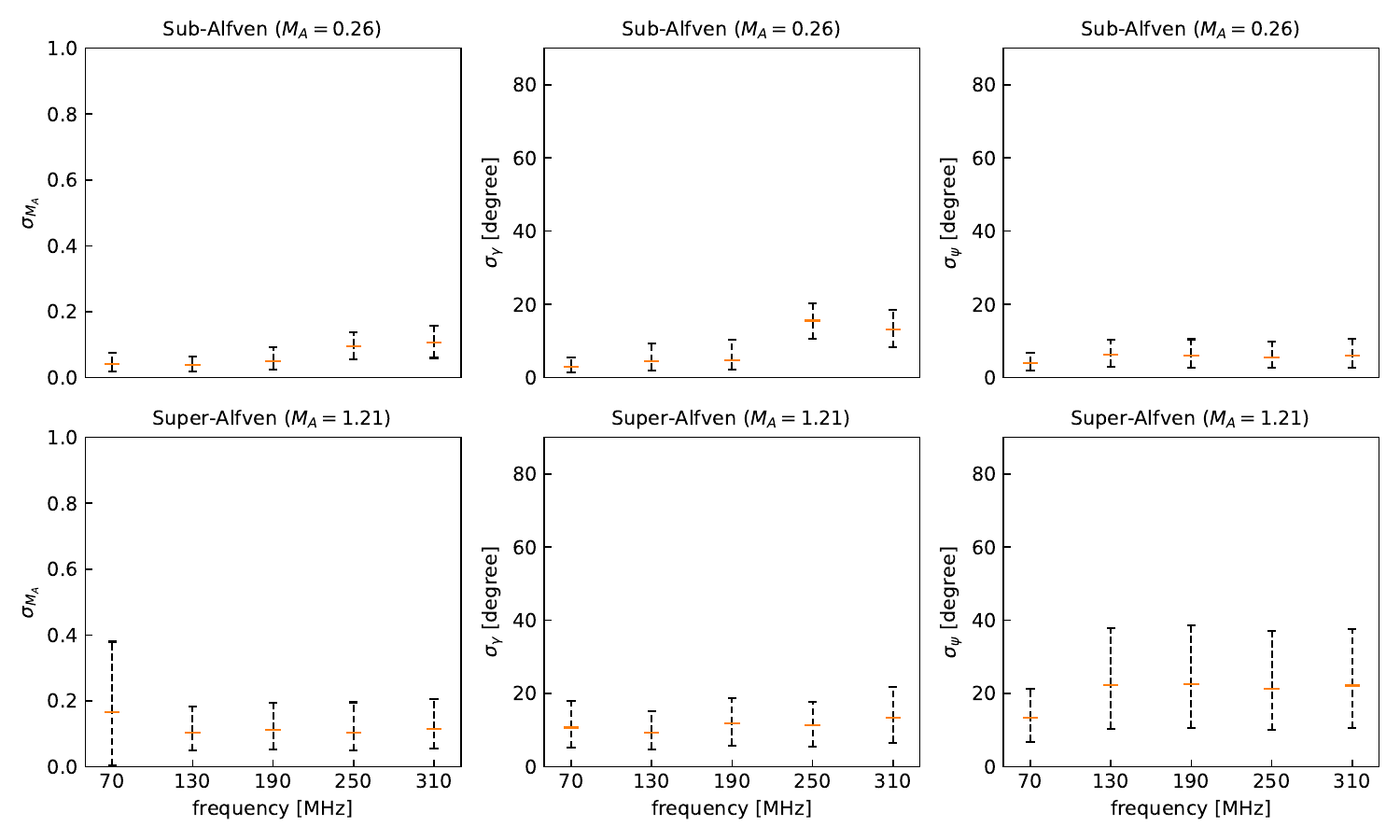}
        \caption{Difference in ViT-estimated $\psi$ (right), $\gamma$ (middle), and $M_A$ (left) and the actual values in the sub-Alfv\'enic ($M_A=0.26$) and super-Alfv\'enic ($M_A=1.21$). The global $\gamma$ is $90^\circ$ for both simulations. The ViT-estimate is performed at 70, 130, 190, 250, and 310 MHz using SPD maps. The upper and lower black lines give ranges of the first (lower) and third quartiles (upper) and the orange line represents the median value. }
    \label{fig:sigma}
\end{figure*}
\section{Results}
\label{sec:result}
\subsection{Fluctuations in SPDs are anisotropic}
We present a numerical example of SPD's anisotropy (see \S~\ref{sec:SPD}) in Fig.~\ref{fig:2Dmap}. The SPD is calculated between 280 MHz and 310 MHz maps. In a strongly magnetized medium (e.g., $M_A=0.50$, $\gamma=90^\circ$), the structures appear more filamentary and anisotropic. As the magnetization weakens (e.g., $M_A=1.21$, $\gamma=90^\circ$), the anisotropy degree decreases. When the magnetic field is not perpendicular to the line of sight (e.g.,  $M_A=1.21$, $\gamma=30^\circ$), the topology of the field lines and the projected synchrotron structures undergo significant changes.

Fig.~\ref{fig:anisotropy_num} presents the anisotropy ratio derived from the SPD maps. These derivatives are calculated across a frequency range from 10 MHz to 310 MHz in increments of 30 MHz. To assess the anisotropy, we calculate the second-order structure function ${\rm SF}_2$ of the $|\frac{d P(\pmb{X},\lambda^2)}{d \lambda^2}|$ map and decompose the structure function into components parallel and perpendicular to the local magnetic fields, using the method proposed in \cite{2000ApJ...539..273C}. The formulation is as follows:
\begin{align}
&\pmb{B}^{\rm loc}_\bot=\frac{1}{2}(\boldsymbol{B}_\bot(\pmb{X}+\pmb{r})+\pmb{B}_\bot(\pmb{X})),\\
&{\rm SF}_2(\pmb{L}_\bot,\pmb{L}_\parallel)=\langle(|\frac{d P(\pmb{X}+\pmb{r},\lambda^2)}{d \lambda^2}|-|\frac{d |P(\pmb{X},\lambda^2)}{d \lambda^2}|)^2\rangle,
\label{eq.sf_loc}
\end{align}
where $\pmb{B}_\bot$ denotes the POS component of the magnetic field, utilized to define the local field direction $\pmb{B}^{\rm loc}_\bot$. Here, the 2D vector $\pmb{L}_\parallel$ is aligned parallel, and $\pmb{L}_\bot$ perpendicular, to $\pmb{B}^{\rm loc}_\bot$, such that $r=\sqrt{L_\parallel^2+L_\bot^2}$ . We determine the $L_\parallel$ value satisfying the condition \footnote{The scale of 20-cells is selected to avoid the numerical dissipation.} ${\rm SF}_2(0,L_\parallel)={\rm SF}_2(20~{\rm cells}, 0)$ to calculate the anisotropy ratio $L_\parallel/L_\bot$.

The upper panel of Fig.~\ref{fig:anisotropy_num} demonstrates  $M_A$ and $\gamma$ can change the anisotropy ratio within the SPD map. When $M_A$ is small and $\gamma$ is large, indicative of a strong magnetic field and minimal projection effect, the structures within the map appear as narrow strips (see Fig.~\ref{fig:2Dmap}), oriented along the POS magnetic fields, leading to a high anisotropy ratio, $L_\parallel/L_\bot$. Conversely, an increase in $M_A$, which signifies a weakening of the magnetic field, results in a decrease in $L_\parallel/L_\bot$. A smaller $\gamma$, implying a magnetic field orientation closer to the LOS, reduces the observed anisotropy due to projection effects. This results in less pronounced elongation along the POS magnetic fields and a lower anisotropic ratio.

Furthermore, the lower panel of Fig.~\ref{fig:anisotropy_num} shows the anisotropy ratio $L_\parallel/L_\bot$ across a frequency range from 40 MHz to 310 MHz. It is observed that $L_\parallel/L_\bot$ consistently exceeds unity, indicating that the structures within the derivative maps predominantly elongate along the magnetic field lines, affirming an anisotropic nature. A strongly magnetized medium also exhibits more pronounced anisotropy. However, the ratio at frequencies below 100 MHz has more variations, due to substantial noise-like contribution from Faraday rotation.

\subsection{Comparison of ViT-estimated 3D magnetic field and actual magnetic field}
Fig.~\ref{fig:3D_sub13} shows a comparative visualization between the actual 3D magnetic fields derived from simulations and those predicted by the trained ViT model. The 3D magnetic field segment in Fig.~\ref{fig:3D_sub13} is constructed from $\psi$ and $\gamma$, with a superimposed color representation indicating $M_A$. The simulation has a global $M_A=0.79$ and $M_s=0.61$ but $\gamma$ is varying along the LOS.  SPD maps at frequencies of 310, 280, and 250 MHz provide insights into the magnetic field structures at distances between [0.49$L_{\rm inj}$, 0.60$L_{\rm inj}$], [0.39$L_{\rm inj}$, 0.49$L_{\rm inj}$], and [0.30$L_{\rm inj}$, 0.39$L_{\rm inj}$], respectively, along the LOS. This enables the reconstruction of the 3D magnetic fields' 3D spatial distribution.

A comparison with the actual magnetic field parameters within the simulation—global $M_A=0.79$ and $M_s=0.61$—reveals a general alignment between the orientations of the ViT-estimates and the actual 3D magnetic fields. Although $\gamma$ is varying, the ViT is able to capture the changes and recover the magnetic fields' variation along the LOS. Furthermore, the actual local sub-field $M_{A,\lambda_1}^{\rm sub}$ is calculated using small-scale local velocity dispersions specific to sub-fields between $L^{\lambda_1}_{\rm eff}$ and $L^{\lambda_2}_{\rm eff}$ along the LOS, resulting in values that are smaller than the global $M_A$. Among our examinations, the median values of ViT-estimated $M_A$ are 0.1 to 0.2 smaller than the actual measurements, indicating an underestimation by the ViT model. Especially, when local $M_A$ variation is significant, the ViT-estimated local $M_A$ could be more significantly underestimated. It means the local cell-level ViT-estimate may not well recover the real values.

\subsection{Uncertainty of the ViT-estimated 3D magnetic fields}
Fig.~\ref{fig:sigma} shows the absolute differences between the ViT-estimated and the actual 3D magnetic field parameters, denoted as $\sigma_\psi$, $\sigma_\gamma$, and $\sigma_{M_A}$. As SPD at high-frequency samples deeper along the LOS, Fig.~\ref{fig:sigma} effectively shows how $\sigma_\psi$, $\sigma_\gamma$, and $\sigma_{M_A}$ change along the LOS. In the sub-Alfv\'enic case ($M_A=0.26$), the median values of $\sigma_\psi$ and $\sigma_\gamma$ predominantly range from 0 to 5$^\circ$ and 0 to 10$^\circ$, respectively, across various frequencies. This reflects a close alignment between the ViT estimates and the actual measurements under sub-Alfv\'enic conditions. The median $\sigma_{M_A}$ is 0.05 - 0.1. As $M_A$ increases to 1.21 in a super-Alfv\'enic condition, $\sigma_\psi$ and $\sigma_\gamma$ expand to ranges of 0 to 20$^\circ$ and 0 to 15$^\circ$, respectively. Despite the broader distribution of $\sigma_{M_A}$ under these conditions, its median value remains consistent at around 0.1. This indicates that the ViT model, compared to earlier CNN models utilized for tracing 3D magnetic fields via synchrotron emission and spectroscopic observations \citep{2024MNRAS.52711240H,2024arXiv240407806H}, is better in estimating $M_A$.

\section{Discussion}
\label{sec:dis}
\subsection{Comparison with other 3D magnetic field tracing approches}
Probing 3D magnetic fields, including the field's POS position angle ($\psi$), inclination angle ($\gamma$) with respect to the LOS, and total magnetization level ($M_A^{-1}$), through machine-learning approaches is rapidly growing. \citet{2024MNRAS.52711240H} proposed the use of Convolutional Neural Networks (CNNs) to extract spatial features, including the elongation direction, anisotropy degree, and morphological curvature, within spectroscopic observation and thereby enable the tracing of 3D magnetic fields. This method is particularly effective for analyzing atomic and molecular emissions within our Galaxy and nearby galaxies. \cite{2024arXiv241111157S}, on the other hand, uses CNN to estimate the sonic Mach number for characterizing turbulence's properties.

\cite{2024arXiv240407806H,2024arXiv241107080Z} expanded the CNN methodology to synchrotron emission observations, which are insensitive to Faraday rotation. Importantly, this study confirmed the robustness of the machine-learning approach even with high-spatial frequencies, making it well-suited for interferometric data that lack single-dish measurements. Furthermore, the CNN model proved capable of extracting useful information from synchrotron emissions and probing 3D magnetic fields, even under conditions of noise with a signal-to-noise ratio greater than unity. It therefore shows great promise for tracing 3D magnetic fields in varied astrophysical settings from ISM, CGM, to ICM.

In this study, we introduce a Vision Transformer (ViT) model designed to estimate the 3D spatial distribution of 3D magnetic fields from synchrotron polarization observations. Especially, we derive the 3D spatial information from the synchrotron polarization derivatives (SPDs): Synchrotron polarization is subject to Faraday decorrelation effects, limiting information to a certain effective path length $L_{\rm eff}$ along the LOS. This means only the regions up to the wave-length dependent $L_{\rm eff}$ contribute to the measured polarization. The difference between synchrotron polarization at two wavelengths, i.e., the SPDs, thus provides insights into the signal’s spatial distribution along the LOS. 

The 3D magnetic field is estimated from the SPDs' anisotropy: (1) the observed SPD structures are predominantly elongated along the local magnetic field lines, making the elongation a direct probe of $\psi$, (2) SPD's anisotropy degree is closely associated with $M_A^{-1}$, and the projection effect induced by $\gamma$, (3) the observed morphological curvature of SPD structures is sensitive to both  $M_A^{-1}$ and $\gamma$. Applying ViT to SPDs allows the extraction of the SPD's features and their correlations with the 3D magnetic field, and thus achieves the reconstruction of the 3D spatial distribution of the 3D magnetic fields. Additionally, compared to the CNN approach, the ViT model provides a more accurate estimation of $M_A^{-1}$. Furthermore, it should be noted that while the anisotropy alone cannot definitively discern whether the magnetic field is oriented toward or away from the observer,  Faraday rotation measurements associated with SPDs potentially can provide the information.

In addition to the machine learning methodology, the Faraday Tomography (FT; \citealt{1966MNRAS.133...67B}) is proposed to trace multiple layers of POS magnetic field orientation along the LOS by analyzing synchrotron polarization at multiple frequencies in the presence of Faraday rotation. It offers high precision but requires extensive frequency coverage and high resolution. \cite{2019MNRAS.485.3499C, 2023MNRAS.519.3736H, 2024ApJ...965..183H} proposed the use of polarized dust emission to constrain the inclination angle averaged along the LOS, based on the depolarization effect. \cite{2024arXiv240714896T} further extend the method to be applicable to stellar polarization. By combining with the information of stats distance, for instance, provided by Gaia \citep{2018A&A...616A...1G}, it is possible to recover the 3D magnetic field's 3D distribution. 

\subsection{Prospects of the ViT-SPD method}

\subsubsection{ISM: 3D Galactic Magnetic Fields}
Understanding the 3D Galactic Magnetic Field (GMF; \citealt{2012ApJ...761L..11J}) is crucial for various astrophysical studies, including tracing the origins of ultra-high energy cosmic rays \citep{2014CRPhy..15..339F,2019JCAP...05..004F} and refining models of Galactic foreground polarization \citep{2015PhRvD..91h1303K,2016A&A...594A..25P}. Previous efforts aimed at modeling the foreground polarization have primarily focused on mapping the POS component of the magnetic field \citep{2019ApJ...887..136C,2020MNRAS.496.2868L,2020ApJ...888...96H},  but neglecting the crucial depolarization factor, magnetic field's inclination angle.

By integrating spectroscopic observations of neutral hydrogen, methods such as velocity channel gradients (VChGs; \citealt{LY18a,2020MNRAS.496.2868L,2020ApJ...888...96H,2023MNRAS.524.2379H}) or CNN-based approaches \citep{2024MNRAS.52711240H}, and the Galactic rotational curve \citep{1985ApJ...295..422C}, it is possible to trace the 3D distribution of the 3D GMF within the Galactic disk. Additionally, employing the ViT method with multi-wavelength synchrotron polarization observations enables the probe of the 3D GMF in the regions of the Galactic halo, thus providing a comprehensive picture of the GMF across the Galaxy.

\subsubsection{ISM \& ICM: Cosmic ray acceleration and transport}
An accurate model of magnetic field properties is crucial for understanding cosmic ray (CR) physics. For instance, the diffusion of CRs is anisotropic and strongly correlated with the medium's magnetization level \citep{2008ApJ...673..942Y, 2013ApJ...779..140X, 2022MNRAS.512.2111H}. Additionally, CR acceleration mechanisms, such as diffusive shock acceleration \citep{1978MNRAS.182..147B, 2000IAUS..195..291A, 2014IJMPD..2330007B, 2022ApJ...925...48X} and turbulent second-order Fermi acceleration \citep{2001MNRAS.320..365B, 2014IJMPD..2330007B}, are instrumental CR acceleration within supernova remnants and galaxy clusters. Both the strength and orientation of magnetic fields play essential roles in determining the efficiency of these acceleration processes. The 3D magnetic field configuration and magnetization level obtained through the ViT-SPDs method can offer critical insights into the acceleration and transport processes of CRs.

\subsubsection{ICM: Magnetic field amplification}
Magnetic field amplification during galaxy cluster mergers plays a pivotal role in understanding cluster evolution. Conventionally, magnetic fields are thought to evolve in tandem with cluster dynamics, where they are stretched, stirred, and amplified by large-scale bulk flows along the merger axis. Subsequent stages involve further amplification by the small-scale turbulent dynamo \citep{1999ApJ...518..594R, 2008ApJ...687..951T, 2018SSRv..214..122D, 2018MNRAS.474.1672V}. This predicted magnetic field topology has recently been confirmed using the synchrotron intensity gradient method as reported by \cite{2024NatCo..15.1006H}. However, more direct assessments, such as magnetic field strength or magnetization distribution within galaxy clusters, are needed. 

Recent studies, such as those by \citet{2024arXiv240407806H}, demonstrate that the machine-learning paradigm of 3D magnetic field tracing is unaffected by the absence of low spatial frequencies in interferometric observations, paving the way for its application with upcoming initiatives like the Square Kilometre Array (SKA; \citealt{2015aska.confE..24B,2015aska.confE..31R}). The ViT-SPD methodology is therefore applicable to interferometric data to trace magnetic field evolution across various redshifts and cluster merger stages, enabling detailed comparisons with predictions of magnetic field amplification.

\subsubsection{Cosmology: constrain axion-like particles}
Axion-like particles (ALPs), predicted by various extensions of the Standard Model, could constitute all or a substantial portion of the cold dark matter \citep{PhysRevLett.38.1440,2010ARNPS..60..405J}. As a class of pseudoscalar particles, ALPs can generically couple to photons. This coupling opens the possibility of oscillations from photons into ALPs in an external magnetic field, which may introduce irregularities in gamma-ray spectra \citep{2013ApJ...772...44W}.  Efforts have been made to explore ALP parameters and the irregularities within the Perseus galaxy cluster \citep{2020PhLB..80235252L}. However, these studies necessitate a detailed magnetic field model. The ViT-SPD methodology could provide essential insights into the 3D magnetic field structure, enabling more stringent constraints on ALP parameters.

\section{Summary}
\label{sec:con}
Probing 3D magnetic fields, including the field's POS position angle ($\psi$), inclination angle ($\gamma$) with respect to the LOS, and total magnetization level ($M_A^{-1}$), presents significant challenges for traditional observational methods. To address these challenges, we introduce the Vision Transformer (ViT) model, leveraging recent advancements in machine learning, to trace 3D magnetic fields using synchrotron polarization derivatives (SPDs). Notably, SPDs provide unique distance information along the LOS, enabling the recovery of the 3D spatial distribution of 3D magnetic fields. We summarize our key findings as follows:
\begin{enumerate}
    \item We demonstrate that the morphological features of SPDs are anisotropic and closely correlated with the 3D magnetic field. The observed SPD structures are predominantly elongated along the local magnetic field lines. The degree of SPD anisotropy is associated with $M_A^{-1}$ and is influenced by the projection effect induced by $\gamma$. Additionally, the observed morphological curvature of SPD structures is sensitive to both $M_A^{-1}$ and $\gamma$.
    \item We developed a ViT model to extract morphological features from SPDs, enabling estimates of the POS magnetic field orientation, inclination angle, and magnetization. Due to the Faraday rotation effect, SPDs provide unique distance information along the LOS, facilitating the recovery of the 3D spatial distribution of the magnetic fields.
    \item The ViT model was trained using synthetic synchrotron polarization data, covering a range of conditions from sub-Alfv\'enic to super-Alfv\'enic regimes. We quantified the uncertainties associated with the model's predictions, finding that the median uncertainties are less than $10^\circ$ for both $\psi$ and $\gamma$, and around 0.05 - 0.1 for $M_A$ under sub-Alfv\'enic conditions. Under super-Alfv\'enic conditions, the uncertainties increased but remained below $20^\circ$ for $\psi$ and $\gamma$, and around $0.1$ for $M_A$.
\end{enumerate}
\begin{acknowledgments}
A.L. acknowledges the support of NSF grants AST 2307840, and ALMA SOSPADA-016. Y.H. acknowledges the support for this work provided by NASA through the NASA Hubble Fellowship grant \# HST-HF2-51557.001 awarded by the Space Telescope Science Institute, which is operated by the Association of Universities for Research in Astronomy, Incorporated, under NASA contract NAS5-26555. This work used SDSC Expanse CPU, NCSA Delta CPU, and NCSA Delta GPU through allocations PHY230032, PHY230033, PHY230091, PHY230105, and PHY240183 from the Advanced Cyberinfrastructure Coordination Ecosystem: Services \& Support (ACCESS) program, which is supported by National Science Foundation grants \#2138259, \#2138286, \#2138307, \#2137603, and \#2138296. 
\end{acknowledgments}

%

\vspace{5mm}


\software{ZEUS-MP/3D code \citep{2006ApJS..165..188H}; Python3 \citep{10.5555/1593511}; TensorFlow \citep{tensorflow2015-whitepaper}}





\bibliography{sample631}{}
\bibliographystyle{aasjournal}



\end{document}